# Tunable Einstein-Bohr recoiling-slit gedankenexperiment at the quantum limit

Yu-Chen Zhang[1,2*], Hao-Wen Cheng[1,2*], Zhao-Qiu Zengxu[1,2*], Zhan Wu[1,2,3], Rui Lin[1,2], Yu-Cheng Duan[1,2,3], Jun Rui[1,2,3], Ming-Cheng Chen[1,2,3], Chao-Yang Lu[1,2,3], Jian-Wei Pan[1,2,3]

[1]Hefei National Research Center for Physical Sciences at the Microscale and School of Physical Sciences, University of Science and Technology of China, Hefei, China.

[2]Shanghai Research Center for Quantum Science and CAS Center for Excellence in Quantum Information and Quantum Physics, University of Science and Technology of China, Shanghai, China.

[3]Hefei National Laboratory, University of Science and Technology of China, Hefei, China.

Corresponding author: cmc@ustc.edu.cn, cylu@ustc.edu.cn, pan@ustc.edu.cn.

## Abstract

In 1927, during the fifth Solvay Conference, Einstein and Bohr described a double-slit interferometer with a "movable slit" that can detect the momentum recoil of one photon[1]. Their debate centered around this gedankenexperiment has provided profound insights into the central concepts of quantum mechanics. Despite many efforts to reproduce this conceptual experiment[2–10], none has realized the original linear optical interferometer faithfully with pure one-photon momentum recoil and full tunability. Here, we report a faithful realization of the Einstein-Bohr interferometer using a single atom in an optical tweezer, which is cooled to the motional ground state in three dimensions[11–14] such that its momentum uncertainty is comparable to that of a single photon. We design an interferometric configuration where the single atom serves as an ultralight, quantum-limit beam-splitter that is momentum-entangled with the input photon. By varying the trap depth of the optical tweezer, we dynamically tune the atom's intrinsic momentum uncertainty, thus enabling the observation of a gradual shift in the visibility of single-photon interference. The interferometer also allows us to distinguish the classical noise caused by atom heating from the quantum-limited noise due to the momentum transfer, illustrating a quantum-to-classical transition[3].

**Main text**

Quantum mechanics embodies inherent uncertainty, encapsulated by two fundamental principles: Heisenberg's uncertainty principle and Bohr's complementarity principle. These concepts challenged our classical intuitions, asserting that for a quantum system, certain pairs of physical properties such as its position and momentum, or wave and particle behaviors, cannot be simultaneously and precisely determined in the same experiment. This inherent uncertainty is not a limitation of our measuring devices, but a fundamental aspect of the nature itself.

These principles sparked a significant debate between Bohr and Einstein. At the 1927 Solvay Conference, Einstein proposed a thought experiment intended to challenge the complementarity principle. Over time, physicists realized that this challenge, far from undermining the quantum theory, profoundly deepened our understanding of quantum measurement and inspired the revolutionary notion of quantum entanglement.

Einstein's thought experiment involved an ingenious modification of the classic double-slit set-up. He suggested a "movable slit" to function as a quantum observer [1]. As shown in Fig. 1a, when a photon interacts with this slit, it may deflect either upwards or downwards, causing the slit to recoil with an equal and opposite momentum. This interaction leads to the entanglement of the photon and the slit, forming a composite quantum state that can be represented as:

$$|\psi(p - \hbar k)\rangle_{\text{slit}} |+\hbar k\rangle_{\text{photon}} + e^{i\phi}|\psi(p + \hbar k)\rangle_{\text{slit}}|-\hbar k\rangle_{\text{photon}}.$$

Here, $|\psi(p)\rangle$ represents the slit's momentum state with uncertainty $\Delta p$, $\hbar k$ is the photon's momentum, and $\phi$ is the phase difference between the two deflected paths.

Einstein argued that by accurately measuring the recoil of the slit, we could determine the photon's trajectory (a particle property) while simultaneously observing interference fringes (a wave property), seemingly conflicting with Bohr's complementarity principle.

We note that for the slit to act as an effective observer, its momentum uncertainty $\Delta p$ must be significantly smaller than the photon's momentum $\hbar k$. Otherwise, the photon's path information would be obscured in the slit's quantum fluctuations. This condition, $\Delta p < \hbar k$, poses a remarkable experimental challenge. For a typical macroscopic slit, say, a 1-g mirror at 100 kHz, its ground-state momentum uncertainty is $\sim 10^{-16}\ kg \cdot m \cdot s^{-1}$, far exceeding the momentum of an optical photon ($\sim 10^{-27}\ kg \cdot m \cdot s^{-1}$).

Despite numerous efforts [2–10] within nearly a century, no experiment has faithfully replicated Einstein's original linear optical interferometer with pure single-photon momentum recoil and a full tunability of the slit's quantum momentum uncertainty. Previous experiments [2–10], while ingenious, did not fully capture the essence of the original proposal. One experiment [15] used photoemission from a diatomic molecule, but destroyed the incoming X-rays in the process. Another work [16] employed a variant of atomic matter-wave interferometry, yet leaked extra which-path information due to unintended classical and quantum correlations. Other variants [3,4] involved additional internal or external degrees of freedom, sidestepping the core challenge of the pure one-photon momentum recoil.

In this work, we realize Einstein's gedankenexperiment using a single Rubidium atom trapped in an optical tweezer [17] as the movable quantum slit (Fig. 1b). The atom is cooled to the three-dimensional motional ground state, which represents the ultimate quantum-limited observer. Its ground-state momentum uncertainty is made comparable to that of a single photon, a regime inaccessible to macroscopic objects. By dynamically tuning the atom's intrinsic momentum uncertainty in the quantum-limited regime, we explore the heart of the Einstein-Bohr debate[1] and observe the transition from wave-like to particle-like behavior of single photons.

The experimental setup, as shown in Fig. 2a, uses an 852 nm laser to form an optical tweezer for trapping a single $^{87}$Rb atom. To make the single atom's intrinsic momentum $\Delta p$ comparable to the single photon's momentum $\hbar k$, we employ three-dimensional Raman sideband cooling [11–14] to prepare the atom into its motional ground state. Here, the most crucial motion is in the axial direction, as the single-photon interference relies heavily on the atom's axial momentum uncertainty in our interferometry set-up (see Fig.

2a). However, the axial trap frequency is ~40 kHz, much lower than the radial trap frequencies of ~300 kHz, which makes the axial cooling more challenging.

To solve this problem, we design a tailored cooling beam geometry, with two Raman beams aligned at 60° to the tweezer axis (Fig. 2a), to maximize the axial coupling while canceling the radial effects[18]. Figure 2b shows the resulting axial sideband spectrum after the Raman cooling, which indicates an axial ground-state occupation of $0.91^{+0.03}_{-0.02}$. The additional data of the radial sideband spectra is shown in Ext. Data Fig. 1, achieving a radial ground-state fidelity of $0.99^{+0.01}_{-0.01}$.

Notably, within the quantum realm of the Einstein-Bohr interferometer, our experiment has the ability to finely adjust the momentum uncertainty of the ground-state atom—the quantum slit. Such tuning is achieved by flexibly changing the tweezer depth. For a ground-state atom in a harmonic potential, the momentum uncertainty is intricately linked to the phonon frequency, which in turn depends on the trap depth, following a fourth-root dependence. In this study, the trap depth is varied from 0.60 mK to 10.49 mK, leading to an optically controllable range of the ground-state atom's momentum uncertainty spanning from 0.78 $\hbar k$ to 1.60 $\hbar k$.

Another critical challenge in faithfully operating this Einstein-Bohr interferometer is mitigating unwanted phase noise from temperature variations and other environmental disturbances between the two optical paths. To this aim, we phase lock the entire system to a 1064-nm reference laser. The laser is split into two beams with a 200-kHz frequency difference to extract the phase offset from the heterodyne signal. Two 5-m optical fibers, wound around a piezoelectric cylinder, enables active stabilization of the path length difference with a large dynamical range of 240 rad. Figure 2c shows the measured residual phase fluctuation of 16.5 mrad rms (corresponding to ~2.8 nm path length fluctuation).

We perform the single-atom recoil experiment by driving the closed cycling transition between $|F = 2, m_F = 2\rangle$ and $|F' = 3, m'_F = +3\rangle$ states using $\sigma^+$-polarized near-resonant light (Fig. 2d). To avoid altering the single atom's internal state, we choose laser powers at a regime of Rayleigh scattering, with saturation parameters of 0.032~0.1. We emphasize that our experimental configuration ensures that the atomic internal state and thermal motion do not leak any which-path information. This makes the atom an ideal quantum beam-splitter envisioned by Einstein. Thus, the first-order interference visibility $V$ of the scattered photons can be written as:

$$V = |\langle\psi(p - \hbar k)|\psi(p + \hbar k)\rangle|^2 = e^{-2\eta^2},$$

which is the overlap of the atom's momentum wavefunctions at two opposite recoil directions [19–21]. The resulting tunable parameter is a ratio between the momentum of the single photon and atom recoil: $\eta = \hbar k/2\Delta p$.

Figure 3a summarizes the key results of the Einstein-Bohr single-atom recoiling-slit gedankenexperiment. The observed single-photon interference raw visibility is plotted (in red circle) as a function of the tweezer trap depth and the tunable parameter $\eta$. Four examples of the interference fringes at different trap depths are shown in Fig. 3b.

As the trap depth increases, the single atom is subject to a stronger spatial confinement. According to the Heisenberg uncertainty principle, the ground-state atom will possess a broader momentum wavefunction. This broader momentum wavefunction leads to a greater degree of overlap after being displaced by the recoil momentum from a single photon, as illustrated in Fig. 1c. The increased level of wavefunction overlap indicates reduced entanglement between the scattered photon and the atom, leading to higher visibility in photon interference. Our observed raw visibilities (red circle) closely align with the trend of the theoretical prediction (black curve), where the movable slit follows the minimum Heisenberg uncertainty principle. Therefore, the observed reduction of the visibility at decreased trap depth in this work is purely from the quantum nature of the interferometer.

However, the values of the observed raw visibilities (red circles) are slightly lower the expected levels in the ideal conditions. This is caused by atom heating, which is due to frequency drift in the 3D Raman sideband cooling, tweezer depth ramping and photon scattering[18]. The effect of atom heating is experimentally calibrated by extracting the residual temperature using scanning Raman spectroscopy in real time. The measured average residual axial phonon number $\bar{n}$ ranges from 0.08 to 0.37 across different points, depicted by purple triangular markers in Fig. 3a, and is considered in the calculation of the blue square dots, which results in a more precise agreement between the experiment data and the refined theory.

The temporal evolution of the Einstein-Bohr interference visibility is illustrated in Fig. 4a, where a reduction in visibility is observed with increased scattering time durations.

Note that the single-photon scattering itself doesn't cause visibility drop[18]. The visibility reduction is because the atom's wavefunction precesses in phase space at the trap frequency (~40 kHz), as depicted in the evolution of the Wigner diagrams in the bottom panel of Fig. 4a. Such a rotation of the atom's wavefunction can increase the position uncertainty $\Delta x$, leading to a larger phase uncertainty. An example is shown for an atom in 10.49-mK trap depth and 6.7-MHz scattering rate, which shows a significant visibility reduction from 84(4)% to 23(4)% over a time scale of 15 µs. Two other examples with controlled parameters at lower tweezer trap depth and lower scattering rates, which results in a much slower visibility reduction, are presented and discussed in[18].

Another source of the visibility reduction is the anti-trapping effect when the atom is in the excited state during the continuous scattering. The anti-trapped excited state causes heating to the atom and spreads over the phase space, which increases the position uncertainty $\Delta x$ and reduces the visibility. The anti-trapping effects, combined with the motional state excitation from scattering, cause the atom's motional state become a thermal state with higher mean phonon number eventually. The red curve in Fig. 4a is numerical calculations taking into account of both effects of the anti-trapping and precession, and is in a good agreement of the observed data.

Finally, we investigate the dependence of the interference visibility as a function of the atom's phonon quantum number $\bar{n}$, which is tuned by controlling the photon scattering time. Here, the single atom having a finite phonon number violates the minimum Heisenberg uncertainty principle. The wavefunction of the atom spreads over the phase space and the interference visibility becomes $V = e^{-2\eta_{eff}^2}$ with $\eta_{eff} = \eta\sqrt{2\bar{n}+1}$ [11]. The observed reduction of the visibility in Fig. 4b at increasing $\bar{n}$ agrees well with the theoretical prediction, which clearly shows the difference between quantum and classical effect on the interference visibility.

## Summary

To summarize, we have faithfully realized Einstein-Bohr linear optical interferometer with a single ground-state atom which acts as a movable slit at one-photon momentum sensitivity. Operating this interferometer at the fundamental level of a single atom and a single photon in pure quantum states, the interference visibility can be continuously tuned by the intrinsic momentum uncertainty of the slit in the quantum limited regime. Using modern language, the Einstein-Bohr interference visibility is determined by the degree of quantum entanglement in the momentum degree of freedom between the photon and the slit. We also investigate the difference between the quantum limit and the classical heating of the atom motional states, which allows us to observe the quantum-to-classical transition.

Future work include implementing state tomography [22–24] of the wave function of the quantum slit, directly probing entanglement between discrete and continuous variables in the photon-slit composite system, and preparing the quantum slit in squeezed state [25]. We also anticipate that gradually increasing the slit's mass in future study will enable us to probe the interplay between decoherence and entanglement, potentially offering new insight in quantum foundations [26,27] and quantum metrology [28].

**Figures**

Figure 1 | The concept of the Einstein-Bohr gedankenexperiment and its realization. **a**, Schematic of single-photon interference with a movable slit. A photon scatters off the

slit, entangling its path with the slit's momentum. **b**, Our experimental realization using a single atom trapped in an optical tweezer and prepared in quantum ground state, which serves as the movable quantum slit. The single atom's momentum uncertainty can be optically tuned to be comparable to a single photon's momentum. A single photon can be Rayleigh scattered either upwards or downwards, which are recombined on a beam splitter. **c**, In this entangled photon-slit system, the slit's momentum states $|\psi(p-\hbar k)\rangle$ and $|\psi(p+\hbar k)\rangle$ correspond to photon paths $|+\hbar k\rangle$ and $|-\hbar k\rangle$, respectively. The interference visibility depends on the overlap of the slit's momentum states.

Figure 2 | Experimental setup. **a**, A single 87Rb atom is trapped in an optical tweezer using a 0.55-NA objective lens. Another objective lens collects scattered photons into a single-mode fiber which are detected by single-photon counting modules (SPCMs). The optical system is phase locked to a 1064-nm reference laser, which is split into two beams by tailor-made dichroic beam displacer (*18*). **b**, Raman sideband spectra demonstrating ground-state cooling of the atom's axial motion. The residual phonon number $\bar{n}$ is calculated by the ratio of the peak of the two sidebands. Error bars represent 1 standard deviation. **c**, The phase stability over a 10-hour measurement, showing a standard deviation of the phase is 16.5 mrad ($\lambda/190$). **d**, Energy levels of the closed recycling transition used for photon scattering.

Figure 3 | Experimental results of the single-photon interference visibility with a tunable and movable quantum-limited slit. **a**, The observed raw interference visibility, plotted in blue square as a function of the trap depth and the normalized recoiling momentum $\eta = \hbar k/2\Delta p$. The black line and red circle are theoretical predictions by assuming a

perfect atomic ground state and considering experimentally calibrated residual phonon number, respectively. Error bars represent 1 standard deviation. **b**, Interference fringes at four different trap depth. The y-axis is the count unbalance of single-photon detections of two arms, calculated by $\frac{C_1-C_2}{C_1+C_2}$ where $C_1(C_2)$ is the photon count of path 1(2). The interference patterns are fitted with sinusoidal function. Error bars are smaller than the marker sizes.

Figure 4 | Reduction of Einstein-Bohr interference visibility due to atom heating and precession. **a**, Measured interference visibilities (circle) as a function of the scattering time and the corresponding theoretical calculation (line). Total scattering time duration is 15 μs. The photon collection time-bin is 1 μs. The scattering rate is 6.7 MHz. The buttom panel shows Wigner diagrams through Monte Carlo simulations illustrating wavefunction evolution in phase space considering scattering, atomic motional state precession and anti-trapping for internal excited state. **b**, Interference visibility versus mean phonon number. The trap depth is 10.49 mK. The data points are the measured interference visibility versus mean phonon number n̄ of the trapped atom. Solid line shows theoretical prediction which follows the equation $V = e^{-2\eta_{\text{eff}}^2}$, where $\eta_{\text{eff}} = \eta\sqrt{2\bar{n}+1}$. Error bars represent 1 standard deviation.

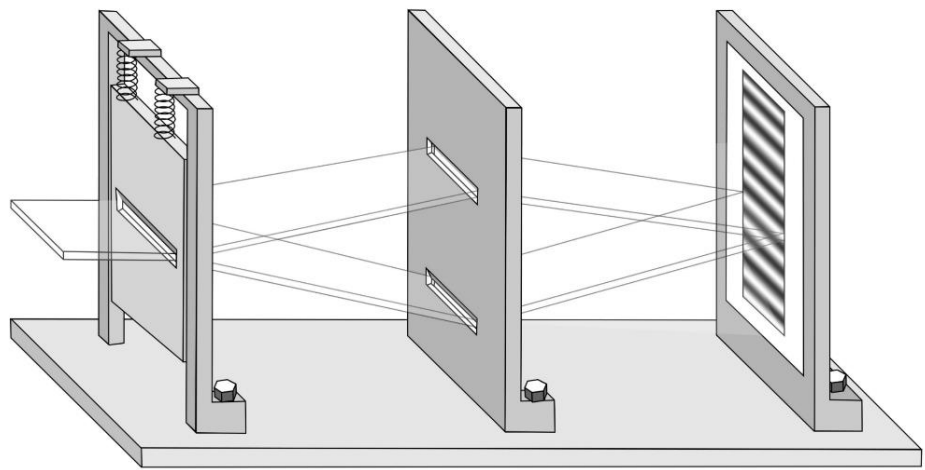
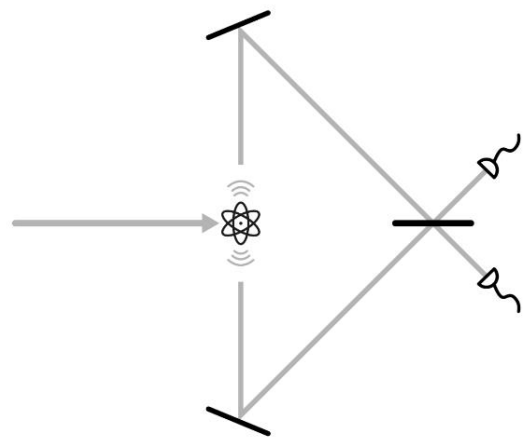
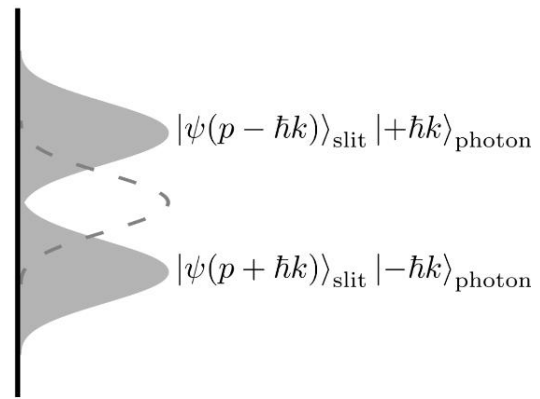

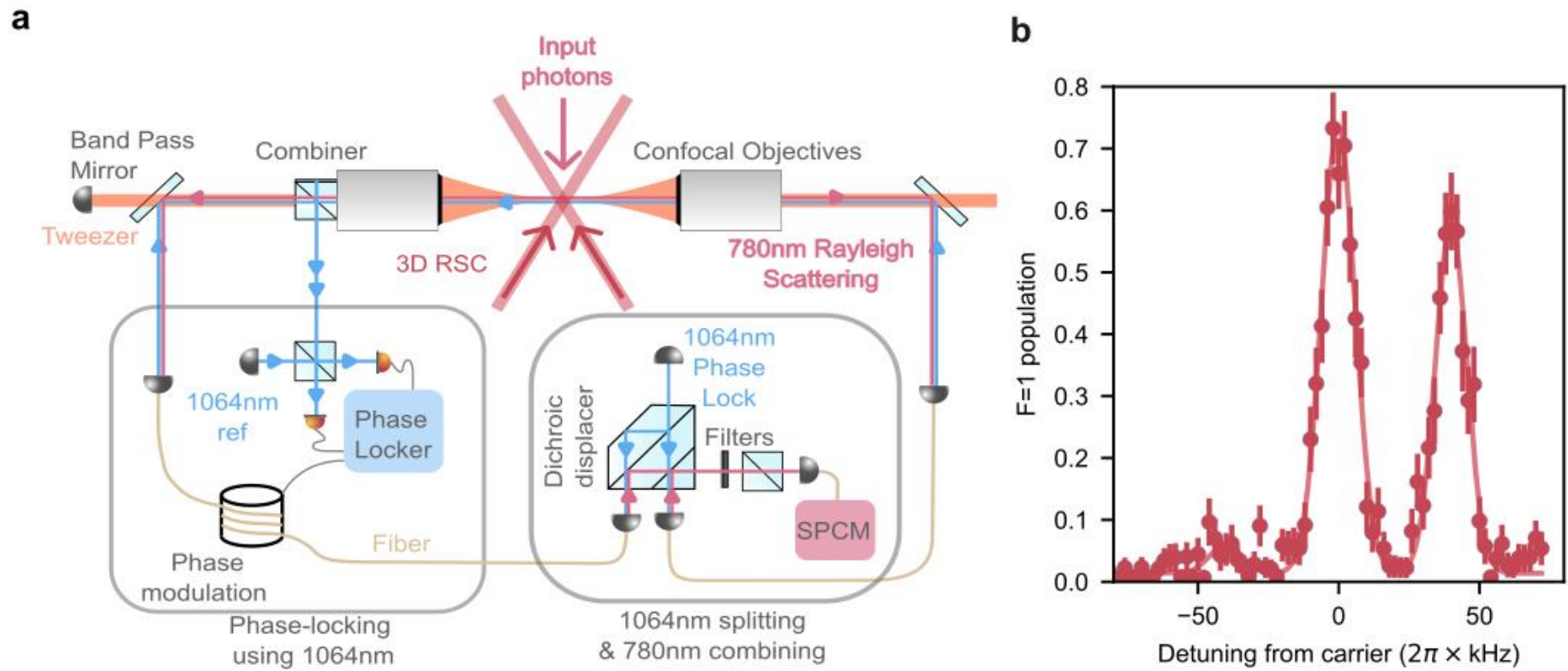
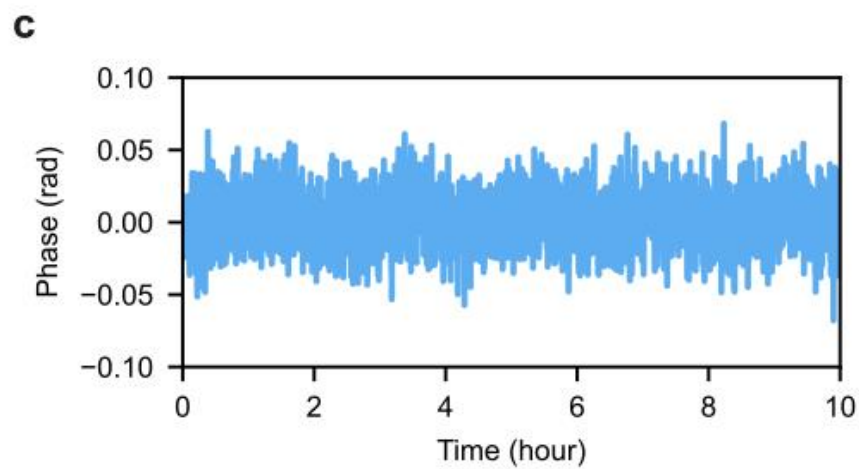
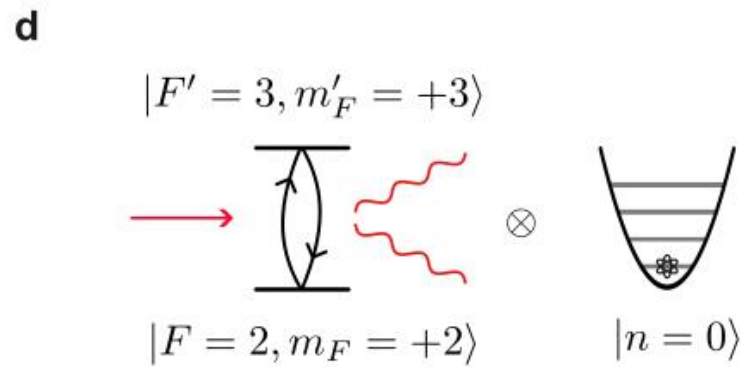

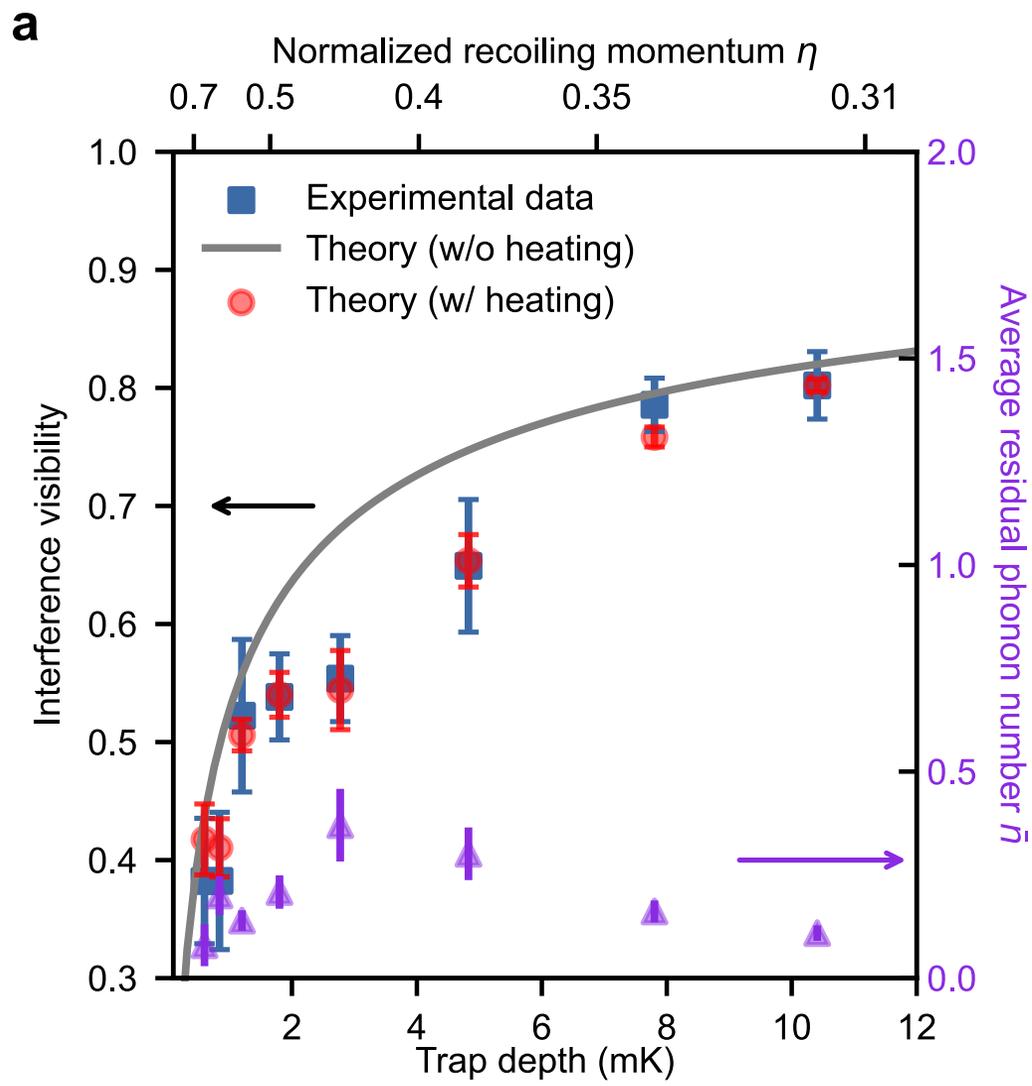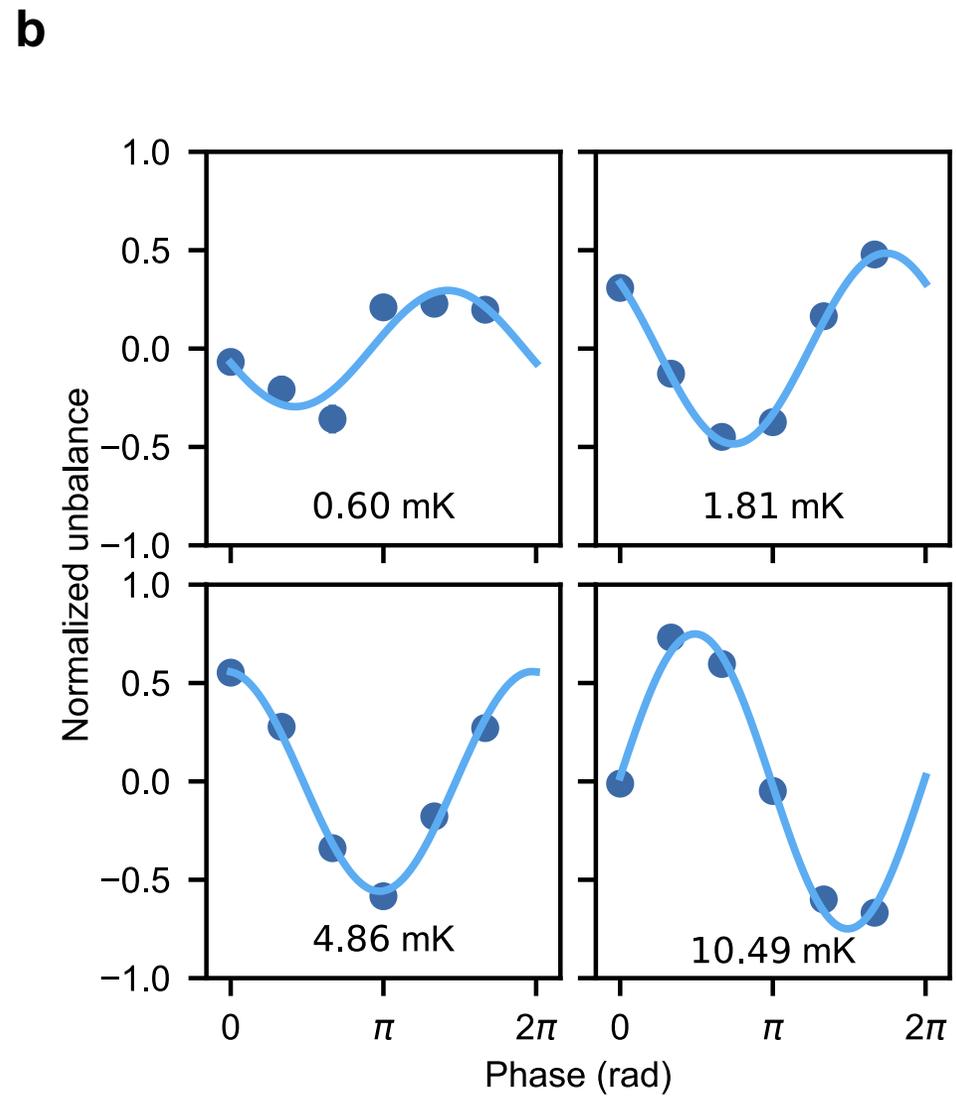

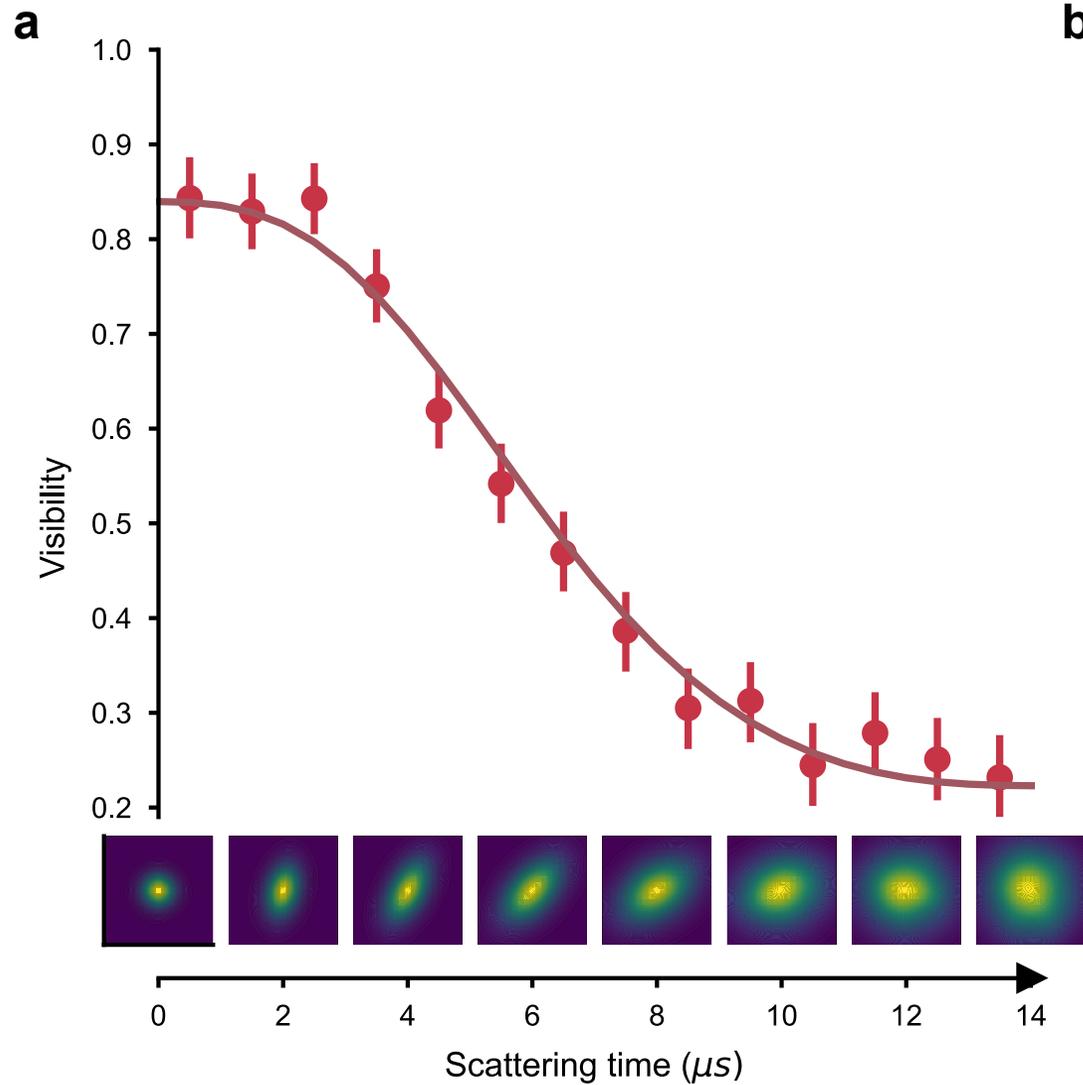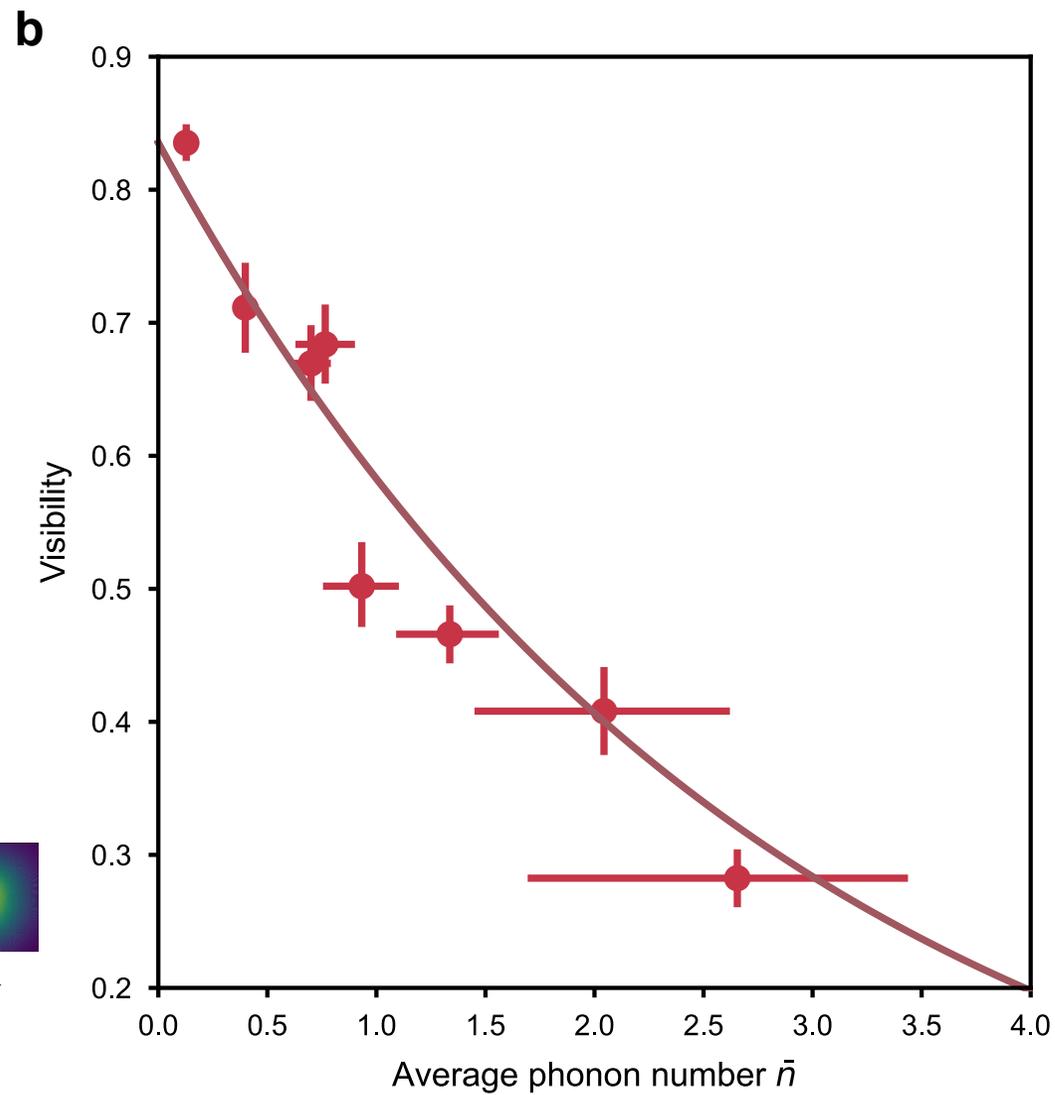